\documentclass[twocolumn,aps,prl,preprintnumbers]{revtex4}
\usepackage{bbm}
\usepackage[latin9]{inputenc}
\usepackage{amsmath}
\usepackage{amssymb}
\usepackage{graphicx}
\usepackage{mathrsfs}
\usepackage{amsfonts}
\usepackage{amsthm}
\usepackage{color}
\usepackage{txfonts}
\usepackage{bm}
\usepackage[colorlinks=true,citecolor=blue,linkcolor=blue,urlcolor=blue,anchorcolor=blue]{hyperref}%
\usepackage{pifont}
\hypersetup{colorlinks=true,citecolor=blue,linkcolor=blue,urlcolor=blue}
\providecommand{\U}[1]{\protect\rule{.1in}{.1in}}
\makeatletter
\@ifundefined{textcolor}{}
{
\definecolor{BLACK}{gray}{0}
\definecolor{WHITE}{gray}{1}
\definecolor{RED}{rgb}{1,0,0}
\definecolor{GREEN}{rgb}{0,1,0}
\definecolor{BLUE}{rgb}{0,0,1}
\definecolor{CYAN}{cmyk}{1,0,0,0}
\definecolor{MAGENTA}{cmyk}{0,1,0,0}
\definecolor{YELLOW}{cmyk}{0,0,1,0}
}

\makeatother
\begin{document}
\title{Strong coupling of chiral magnons in altermagnets}
\author{Zhejunyu Jin$^1$}
\author{Tianci Gong$^1$}
\author{Jie Liu$^1$}
\author{Huanhuan Yang$^1$}
\author{Zhaozhuo Zeng$^1$}
\author{Yunshan Cao$^1$}
\author{Peng Yan$^{1,2}$}
\email[Contact author: ]{yan@uestc.edu.cn}
\affiliation{$^1$School of Physics and State Key Laboratory of Electronic Thin Films and Integrated Devices, University of
Electronic Science and Technology of China, Chengdu 610054, China\\
$^2$Institute of Fundamental and Frontier Sciences, Key Laboratory of Quantum Physics and Photonic Quantum Information of the Ministry of Education, University of Electronic Science and Technology of China, Chengdu 611731, China}

\begin{abstract}
Altermagnets recently are identified as a new class of magnets that break the time-reversal symmetry without exhibiting net magnetization. The role of the dipole-dipole interaction (DDI) on their dynamical properties however is yet to be addressed. In this work, we show that the DDI can induce the strong coupling between exchange magnons with opposite chiralities in altermagnets, manifesting as a significant level repulsion in the magnon spectrum. Crucially, the predicted magnon-magnon coupling is highly anisotropic, and observable in practical experiments. These exotic features are absent in conventional antiferromagnets. Our findings open a new pathway for quantum magnonic information processing based on altermagnetism.
\end{abstract}

\maketitle

\textit{Introduction---}Magnons (quanta of spin waves) have been intensively studied for wave-based sensing and computing concepts, due to their long lifetime and high tunability \cite{Kruglyak2010,Chumak2015,Pirro2021}. The compatibility between magnons and diverse quantum platforms such as qubits \cite{Tabuchi2015,Mq2022,Mq2023}, phonons \cite{Agrawal2013,Streib2019,Bozhko2020}, and photons \cite{Bai2011,Braggio2016,Harder2018} further amplifies the advantages of magnons as an ideal carrier for quantum information processing, constituting quantum magnonics \cite{Yuan2022}. Coherent information transfer between two magnonic systems demands an effective magnon-magnon coupling that is usually generated by the dipole-dipole interaction (DDI) \cite{Shiota2020}, interlayer exchange \cite{Chen2018,Ndiaye2017,Sklenar2021}, and in-plane anisotropy \cite{Liensberger2019}, etc. Meanwhile, it can also be realized by indirect approaches, such as cavity photons \cite{Soykal2010,Yuan2017,Lambert2016,Rameshti2018,Grigoryan2019,Nair2022, Zhang2023,Johansen2018}. Despite significant advances, previous studies focused on long-wavelength magnons in isotropic systems, where the coupling is independent on the direction of magnon propagation \cite{He2021}. Achieving anisotropic coupling between short-wavelength magnons should be a crucial step for implementing directional control in magnon-based circuits, but it remains a major challenge in the community.

It has been known that the DDI plays a prominent role in shaping magnon dispersions and stabilizing spin textures \cite{Serga2010,Montoya2017} in ferromagnets. In contrast, it is negligible in antiferromagnets due to the near-perfect cancellation of magnetization between sublattices. While an external magnetic field can induce a net magnetization, it breaks the degeneracy between magnon branches of opposite chiralities. As a result, the DDI only introduces a weak anisotropy, slightly shifting the magnon levels without generating mode coupling.
\begin{figure}[htbp]
  \centering
  \includegraphics[width=0.48\textwidth]{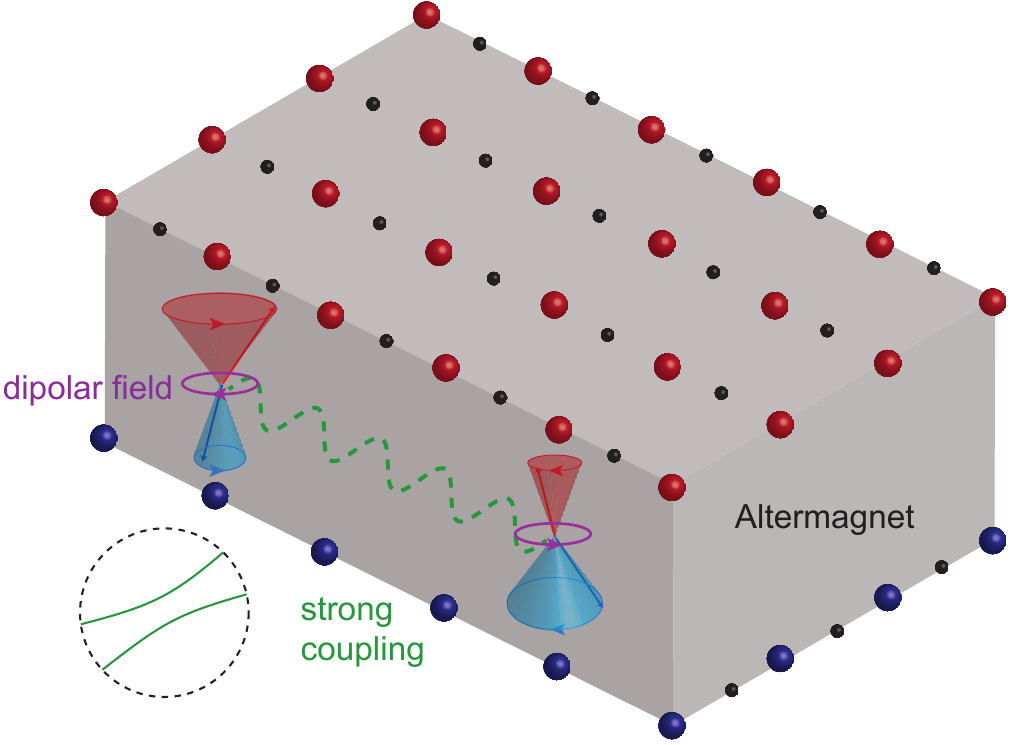}\\
 \caption{Schematics of the dipolar-interaction induced strong coupling of magnons in altermagnets, manifesting as a level repulsion (solid green curves). The effective coupling (dashed green wavy line) between two magnons with opposite chiralities is mediated by the dipolar field (purple arrow).}\label{fig1}
\end{figure}

A newly discovered class of magnetism, dubbed altermagnetism \cite{SongNRM2025}, offers a promising alternative. Altermagnets exhibit a novel symmetry involving combined real- and spin-space operations, leading to properties intermediate between ferromagnets and antiferromagnets \cite{Hayami2019,Hayami2020,Smejkal1,Smejkal2,Smejkal3,Mazin2023PRB,Turek2022,Feng2022,Ouassou2023,SZhang2023,Zhou2023,Hariki2023,Sun2023,Bai2023,Bialek2023,Ghorashi2023,yutao2023,Mazin2023}. Unlike collinear antiferromagnets, altermagnets can exhibit intrinsic spin splitting even without external magnetic fields, due to the non-centrosymmetric symmetry operations that relate the two magnetic sublattices. This symmetry further allows magnons with opposite chiralities to be connected via point group rotations, resulting in protected nodal lines in the magnon band structure. When a weak external magnetic field is applied, the degenerate feature can persist\rule[2pt]{0.4cm}{0.06em}as long as the Zeeman energy remains below the characteristic magnon bandwidth\rule[2pt]{0.4cm}{0.06em}preserving level crossings at finite wavevectors (see below). This resilience to field-induced splitting offers an ideal platform for exploring DDI-driven physics. In particular, the DDI can now act in a non-perturbative manner, i.e., lifting degeneracies and inducing strong coupling between magnons of opposite chiralities, which is not accessible in conventional antiferromagnets \cite{Sheng2020,Jie2023}. Moreover, the intrinsic anisotropy in the exchange and crystal structure of altermagnets gives rise to directionally dependent magnon dispersions, further enhancing the prospects for anisotropic magnon-magnon coupling.

In this work, we explore the role of the DDI in the magnon spectrum of altermagnets. Without loss of generality, we take the $d$-wave altermagnet as the model system. We first find that a magnetic field leads to an upward (downward) shift of the magnon branch with right (left)-handedness, preserving the level crossing at a finite wavevector. Then, we show that the DDI breaks the conservation of spin angular momentum conservation, resulting in a strong coupling between exchange magnons with opposite chiraliries, manifesting as a level repulsion, as shown in Fig. \ref{fig1}. We observe that the magnon-magnon coupling strongly depends on the magnon propagation direction due to the anisotropic nature of the exchange interaction. Analytical results are verified by full micromagnetic simulations. Our findings open the door for exploring quantum magnonics in the exotic platform of altermagnetism.

\begin{figure}[htbp]
  \centering
  \includegraphics[width=0.48\textwidth]{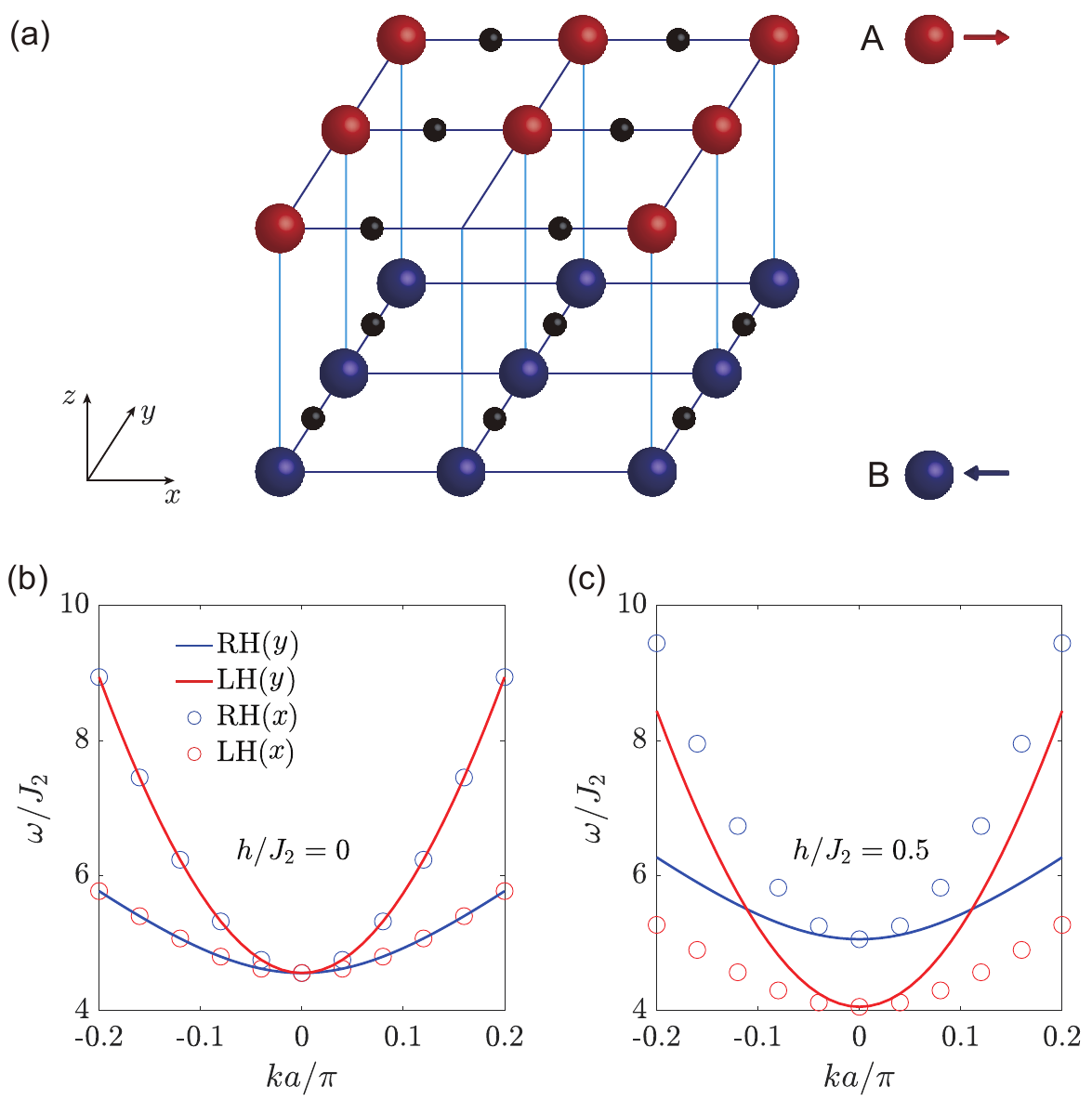}\\
 \caption{(a) Illustration of a two-sublattice altermagnet. The black spheres correspond to nonmagnetic atoms. Magnon dispersion for $h/J_2=0$ (b) and $0.5$ (c). Curves and circles represent the magnons propagating along $y$ and $x$ direction, respectively. The handedness/chirality of magnons is distinguished by contrast color (blue and red).}\label{fig2}
\end{figure}

\textit{Level-crossing without DDI---}We consider a bilayer altermagnet modeled by the following Hamiltonian
\begin{equation}\label{Eq1}
\begin{aligned}
{\mathcal H}_{\text{alter}}=&-\sum_{i,j}\big[J_{1}{\bf S}_{i,j}^A\cdot{\bf S}_{i+1,j}^A+J_{2}{\bf S}_{i,j}^B\cdot{\bf S}_{i+1,j}^B\\
&+J_{2}{\bf S}_{i,j}^A\cdot{\bf S}_{i,j+1}^A+J_{1}{\bf S}_{i,j}^B\cdot{\bf S}_{i,j+1}^B+J_3{\bf S}_{i,j}^A\cdot{\bf S}_{i,j}^B\\
&+{\bf h}\cdot({\bf S}_{i,j}^A+{\bf S}_{i,j}^B)+K({\bf S}_{i,j}^A\cdot{\bf x})^2+K({\bf S}_{i,j}^B\cdot{\bf x})^2\big],\\
\end{aligned}
\end{equation}
where $J_{1,2}>0$  represents the intralayer ferromagnetic exchange coupling strength, $J_3<0$ is the interlayer antiferromagnetic exchange coupling coefficient, ${\bf h}$ and $K$ denote the external magnetic field and the magnetic anisotropy constant, respectively, ${\bf S}_{i,j}^A$ and ${\bf S}_{i,j}^B$ are the spin vectors on sites $(i,j)$ of sublattices $A$ and $B$, respectively. Figure \ref{fig2}(a) shows the crystal structure of a two-sublattice altermagnet. Under a combined operation of two-fold spin-space rotation $\mathcal{C}_{s,2}$ (${\bf S}^A\rightarrow -{\bf S}^A$, ${\bf S}^B\rightarrow -{\bf S}^B$), four-fold crystallographic-space rotation $\mathcal{C}_4$ $(i\rightarrow j$ and $J_1\rightarrow J_2)$, and an additional glide operation, we find that Hamiltonian \eqref{Eq1} is invariant and it thus respects the symmetry of $d$-wave altermagnets \cite{Smejkal1}. Notably, altermagnet can be seen as an antiferromagnet with anisotropic exchange coupling, and this similarity provides a potential for their mutual transformations \cite{Mazin2023}. We then obtain the magnon dispersion (see Sec. I in Supplemental Material \cite{SM})
\begin{equation}\label{Eq2}
\begin{aligned}
\omega_{{\bf k},\pm}&=\pm c_1({\bf k})+c_2({\bf k}),
\end{aligned}
\end{equation}with $c_1({\bf k})=S\Big \{h/S+(J_1-J_2)\big[\cos(k_ya)-\cos(k_xa)\big]\Big \}$ and $c_2({\bf k})=S\Big \{2K-(J_2+J_1)\big[\cos(k_xa)+\cos(k_ya)-2\big]\Big \}^{1/2}\Big \{2K-2J_3-(J_1+J_2)\big[\cos(k_xa)+\cos(k_ya)-2\big]\Big \}^{1/2}$. Here, $a$ is the lattice constant and $\pm$ corresponds to right-handed (RH) and left-handed (LH) modes, with respect to the $x$ axis, respectively, and $S$ represents the spin length. In what follows, we use the following parameters to calculate the spectrum: $J_1= 6.53J_2$, $J_3 = -3.22J_2$, $K = 0.6J_2$, and $S=1.5$ \cite{yutao2023}, if not stated otherwise. Different from the case in conventional antiferromagnets, the degeneracy of RH and LH magnons is broken except the $\Gamma$ point ($k=|{\bf k}|=0$), exhibiting contrasting group velocities of two branches, as shown in Fig. \ref{fig2}(b). Additionally, it is found that the spectrum of RH magnons propagating along the $x$ $(y)$-direction is identical to that of LH magnons propagating along the $y$ $(x)$ direction. When a in-plane magnetic field along $x$-direction is applied, the energy degeneracy at $k=|{\bf k}|=0$ is removed. However, we observe that, for magnons propagating along $y$-direction, the branch with a lower group-velocity shifts upward, while the one with a higher group-velocity shifts downward. This subsequently preserves the level-crossing at a finite wave number [see red and blue curves in Fig. \ref{fig2}(c)]. Such feature does not exist for magnons propagating along the $x$ direction, which is an indication of anisotropic level-crossing and shall be discussed below. It is noted that this persistent level-crossing is a generic feature for altermagnetic magnons, regardless of the concrete model (see Sec. II \cite{SM}).

\textit{Level repulsion with DDI---}A level-crossing usually means the absence of coupling. However, the DDI can mix magnon's spin and orbit degrees of freedom and consequently breaks the spin conservation. In our model, the dipolar interaction reads
\begin{equation}\label{Eq3}
\begin{aligned}
{\mathcal H}_{\text{DDI}}=&\kappa \sum_{(l,i,j) \ne (l',i',j')}  \Big [\frac{{\bf S}_{i,j}^{l} \cdot {\bf S}_{i',j'}^{l'}}{{| {{{\bf R}_{ij,i'j'}^{l,l'}}} |}^3}-3\frac{({\bf S}_{i,j}^{l} \cdot{\bf R}_{ij,i'j'}^{l,l'} )({\bf S}_{i',j'}^{l'}\cdot {\bf R}_{ij,i'j'}^{l.l'})}{{| {{{\bf R}_{ij,i'j'}^{l,l'}}} |}^5}\Big ],
\end{aligned}
\end{equation}
\begin{figure}[htbp]
  \centering
  \includegraphics[width=0.48\textwidth]{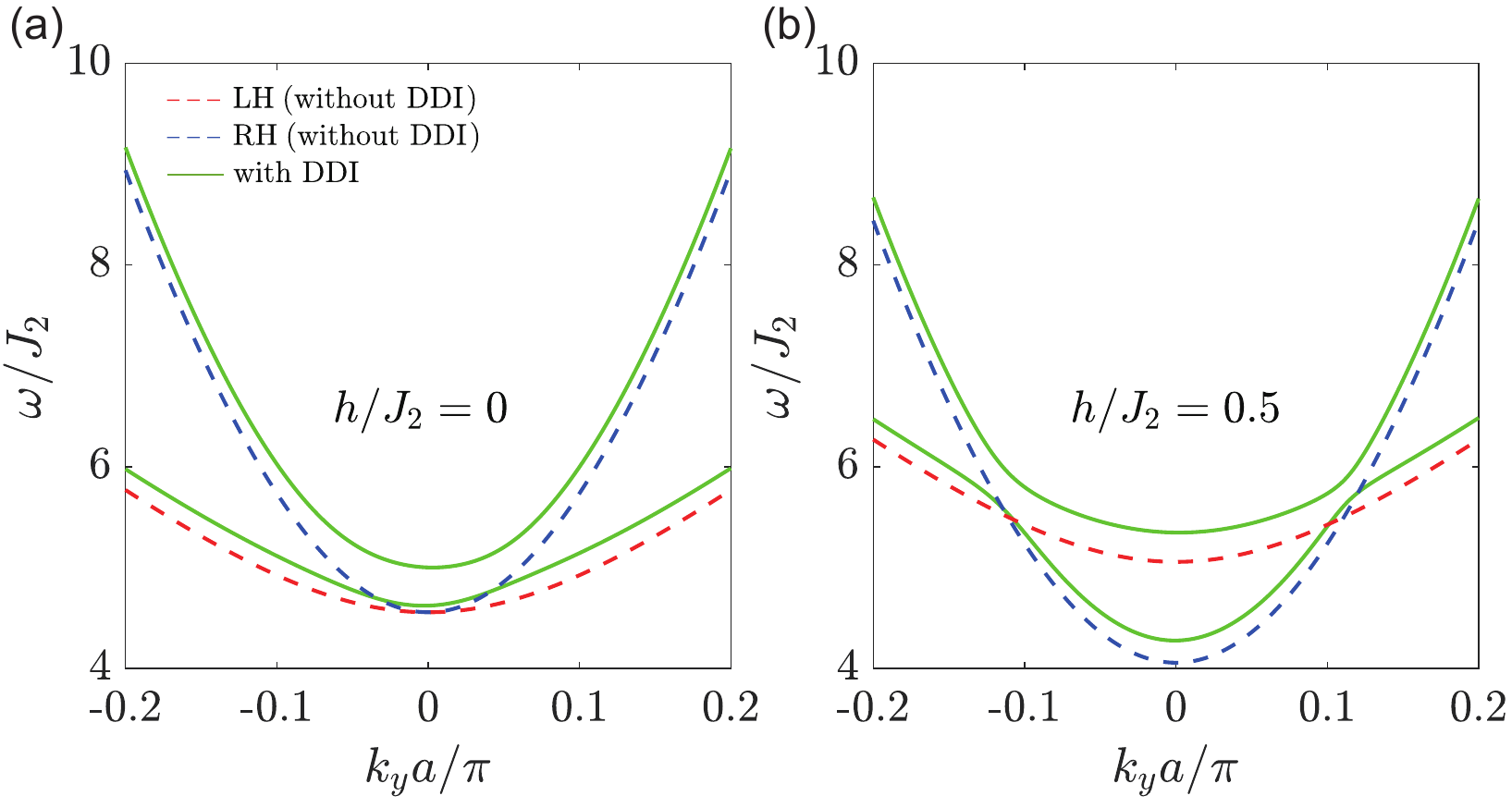}\\
 \caption{Magnon dispersion for $h/J_2=0$ (a) and $0.5$ (b). The solid (dashed) lines stand for the results with (without)
the DDI. }\label{fig3}
\end{figure}where $l$ and $l'$ represent the layer indexes, $\kappa = \mu_0(g\mu_B)^2/2$ with $\mu_0$ the vacuum permeability, $g$ the Lander factor, and $\mu_B$ representing the Bohr magneton, and ${\bf R}_{ij,i'j'}^{l,l'}={\bf R}_{ij}^{l}-{\bf R}_{i'j'}^{l'}$ is the distance between two spin sites. In Eq. \eqref{Eq3}, we have included both intra- and inter-layer DDIs.
By performing the Holstein-Primakoff (HP) transformation \cite{HP}: $S^{A,+}=\sqrt{2S}a,S^{B,+}=\sqrt{2S}b^\dag, S^{A,-}=\sqrt{2S}a^\dag,S^{B,-}=\sqrt{2S}b, S^A_x=S-a^\dag a,$ and $S^B_x=b^\dag b-S$, where the spin operators $S^{m,\pm}=S^m_z\pm iS^m_y$ with $m=A,B$, and $a\ (b)$ and $a^\dag\ (b^\dag)$ are the magnon annihilation and creation operators for sublattice $A\ (B)$, respectively, we can express the altermagnetic Hamiltonian in the momentum space under the basis $(a_{\bf k}, b_{\bf k}, a_{- \bf k}^{\dag}, b_{- \bf k}^{\dag} )^\text{T}$
\begin{equation}\label{Eq4}
\begin{aligned}
{\mathcal H}_{\text{alter}}={\mathcal H}_{\text{alter},\bf k}^{0}+\mathcal{H}_{\text{DDI},\bf k},
\end{aligned}
\end{equation}
with the DDI-free Hamiltonian 
\begin{equation}\label{Eq5}
\mathcal{H}_{\text{alter},\bf k}^0=\left(
    \begin{array}{cccc}
      H_{11} & 0 & 0 & SJ_3\\
      0 & H_{22} & SJ_3 & 0\\
      0 & SJ_3 & H_{11} & 0\\
      SJ_3 & 0 &  0 & H_{22}\\
    \end{array}
  \right),
\end{equation}
where $H_{11}=h+2KS+2S\bigg\{J_1\big[\cos(k_x a)-1\big]+J_2\big[\cos(k_y a)-1\big]\bigg\}+SJ_3$, $H_{22}=-h+2KS+2S\bigg\{J_2\big[\cos(k_xa)-1\big]+J_1\big[\cos(k_ya)-1\big]\bigg\}+SJ_3$, and the DDI part
\begin{equation}\label{Eq6}
\mathcal{H}_{\text{DDI},\bf k}=\left(
    \begin{array}{cccc}
      C_{11,{\bf k}} & C_{12,{\bf k}} & C_{13,{\bf k}} & C_{14,{\bf k}}\\
      C_{12,{\bf k}} & C_{22,{\bf k}} & C_{14,{-\bf k}}^* & C_{24,{\bf k}}\\
      C_{13,{\bf k}}^* & C_{14,{-\bf k}} & C_{11,{\bf k}} & C_{12,{-\bf k}}\\
      C_{14,{\bf k}}^* & C_{24,{\bf k}}^* &  C_{12,{-\bf k}} & C_{22,{\bf k}}\\
    \end{array}
  \right).
\end{equation}See Sec. III \cite{SM} for the detailed expression of $C_{ij,{\bf k}}$. To obtain the magnon spectrum, we use the paraunitary transformation to diagonalize the above matrix.

Figures \ref{fig3}(a) and (b) show the magnon dispersion without and with the external magnetic field, respectively. Clearly, the DDI opens a gap at the intersection region, which manifests as a level repulsion. Interestingly, we note that the magnon-magnon coupling is asymmetric along the $+y$ and $-y$ directions.
\begin{figure}[htbp]
  \centering
  \includegraphics[width=0.48\textwidth]{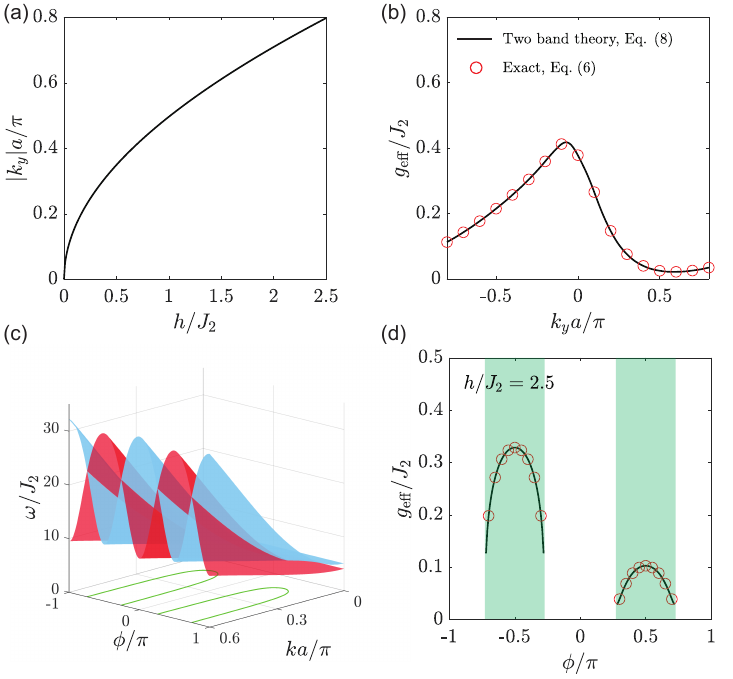}\\
 \caption{(a) Field-dependence of the level-crossing point. (b) DDI-induced $g_{\rm{eff}}$ as a function of $k_y$. 
 (c) Magnon spectrum (blue and red surfaces) as a function of $k$ and $\phi$, with the green curve being the projection of the intersection. (d) Anisotropic magnon-magnon coupling. Green area labels the anticrossing region. Black curves and red circles in (b) and (d) represent the analytical and numerical results, respectively. In (c) and (d), we set $h/J_2=2.5$.}\label{fig4}
\end{figure}
To provide a quantitative understanding of these features, we project the $4\times4$ matrix \eqref{Eq6} into a reduced subspace under the basis consisting of the eigenstates of magnons with opposite handedness $\alpha_{\bf k}$ and $\beta_{\bf k}$. Here $\alpha_{\bf k}=u_{\bf k}a_{\bf k}-v_{\bf k}b_{\bf k}^\dag$, $\beta_{\bf k}=u_{\bf k}b_{\bf k}-v_{\bf k}a_{\bf k}^\dag$ with $u_{\bf k}=\sqrt{(\Delta_{\bf k}+1)/2}$, $v_{\bf k}=-\sqrt{(\Delta_{\bf k}-1)/2}$, and
\begin{equation}\label{Eq7}
\frac{1}{\Delta_{\bf k}}=\sqrt{{1-\Big\{\frac{J_3}{(J_1+J_2)[\cos(k_xa)+\cos(k_ya)-2]+J_3-2K}\Big\}^2}}.
\end{equation}
The effective two-band Hamiltonian under such approximation is expressed as
\begin{equation}\label{Eq8}
\mathcal{H}_{\text{DDI},\bf k}^{\rm eff}=\left(
    \begin{array}{cc}
      D_{11,{\bf k}} & D_{12,{\bf k}}\\
      D^{*}_{12,{\bf k}} & D_{22,{\bf k}}\\
    \end{array}
  \right),
\end{equation}
where
\begin{equation}\label{Eq9}
\begin{aligned}
D_{11,{\bf k}}&=u_{\bf k}(C_{11,{\bf k}}u_{\bf k}+C_{14,{\bf k}}^*v_{\bf k})+v_{\bf k}(C_{14,{\bf k}}u_{\bf k}+C_{22,{\bf k}}v_{\bf k}),\\
D_{12,{\bf k}}&=u_{\bf k}(C_{12,{\bf k}}u_{\bf k}+C_{24,{\bf k}}^*v_{\bf k})+v_{\bf k}(C_{13,{\bf k}}u_{\bf k}+C_{12,{-\bf k}}v_{\bf k}),\\
D_{22,{\bf k}}&=u_{\bf k}(C_{22,{\bf k}}u_{\bf k}+C_{23,{\bf k}}^*v_{\bf k})+v_{\bf k}(C_{23,{\bf k}}u_{\bf k}+C_{11,{\bf k}}v_{\bf k}).\\
\end{aligned}
\end{equation}
Equation \eqref{Eq8} shows that the DDI breaks the degeneracy of magnons with opposite chiralities, and the resulting effective magnon-magnon coupling is $g_{\rm{eff}}=2|D_{12,{\bf k}}|$. The matrix elements $C_{12,{\bf k}}$ and/or $D_{12,{\bf k}}$ are asymmetric along the $\pm y$ directions (see Sec. III \cite{SM}), a consequence of the Damon-Eshbach geometry \cite{Gallardo2019}, generating an asymmetric magnon-magnon coupling. Moreover, we observe that the effective coupling is wavevector-dependent, and the Hamiltonian can be recast as $\mathcal{H}_{\text{DDI},\bf k}^{\rm eff}=\frac{1}{2}(D_{11,{\bf k}}+D_{22,{\bf k}}){\bf I}+{\bf f}({\bf k})\cdot{\bm \sigma}$, with $f_x({\bf k})={\rm Re}(D_{12,{\bf k}})$, $f_y({\bf k})={\rm Im}(D_{12,{\bf k}})$, $f_z({\bf k})=\frac{1}{2}(D_{11,{\bf k}}-D_{22,{\bf k}})$, and $\bm \sigma$ being the Pauli matrices, which manifests as an effective spin-orbit coupling for magnons.

Figure \ref{fig4}(a) shows the field-dependence of the level-crossing point. It is found that the critical wavevector is pushed to the short-wavelength regime, due to the increased Zeeman splitting between two magnon modes. Figure \ref{fig4}(b) plots the calculated coupling strength as a function of the wavevector at the level-crossing point, based on the four-band (curves) and two-band (circles) Hamltonians. Two approaches give consistent results across the entire Brillouin zone. In the above calculations, we focus on the magnon propagation along the $y$ direction. However, the anisotropic nature of the exchange interaction in altermagnets is expected to generate anisotropic magnon dispersion. We then examine if the magnon-magnon coupling depends on the propagation direction of magnons. To this end, we express the magnon wavevector as $\mathbf{k}=k(\cos\phi \hat{x}+\sin\phi \hat{y})$ with $\phi$ being the polar angle. Then, we derive the relation between the wavevector $k_c$ at the level-crossing point and the angle $\phi$ as
\begin{equation}\label{Eq10}
\begin{aligned}
\cos(k_ca\cos\phi)-\cos(k_ca\sin\phi)&=\frac{h}{S(J_{1}-J_{2})},
\end{aligned}
\end{equation}which can only be solved numerically. Figure \ref{fig4}(c) shows the magnon spectrum as a function of $k$ and $\phi$. We observe that $k_c$ (the green curve) is symmetric about $\phi = \pm\pi/2$ due to the mirror symmetry of the crystal with respect to the $x-z$ plane, while it takes the minimum at $\phi=\pi/2$ or $-\pi/2$ because the difference in the group velocities of magnons with opposite chiralities reaches the maximum in such cases. When $\phi$ deviates from these two values, $k_c$ substantially increases. In addition, it is noted that Eq. \eqref{Eq10} has real solutions only when the magnon propagation angle $\phi$ falls into two windows $[\pi/2-\phi_0,\pi/2+\phi_0]$ and $[-\pi/2-\phi_0,-\pi/2+\phi_0]$ with $\phi_0\approx\sqrt{4+2h/[S(J_1-J_2)]}/\pi$. As the in-plane magnetic field increases, the level-crossing point rapidly moves away from the origin (not shown). A direct consequence of the $\phi$-dependence of $k_c$ is the anisotropy of the effective coupling $g_{\rm{eff}}$. 

Figure \ref{fig4}(d) displays the magnon-magnon coupling as a function of $\phi$. Numerical calculations (symbols) are in full agreement with the predictions of the two-band model (curves). Notably, at $h/J_2=0.3$, the effective magnon-magnon coupling reaches $g_{\rm eff}\sim 0.1 \omega_{{\bf k},\pm}$, indicating a strong magnon-magnon coupling originating from the DDI.

\begin{figure}[htbp]
  \centering
  \includegraphics[width=0.48\textwidth]{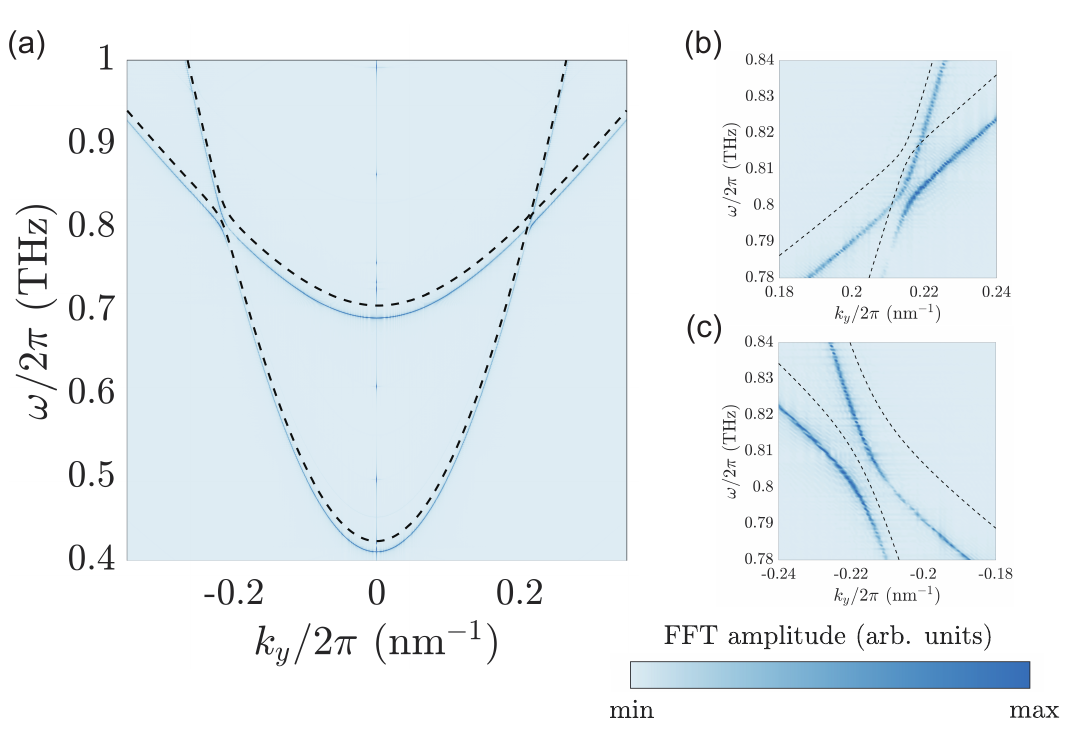}\\
 \caption{(a) The dispersion relation of altermagnetic magnons obtained by the fast Fourier transformation (FFT) of the dynamical magnetization from full micromagnetic simulations. The dashed black curves represent the analytical results from the linear spin-wave theory. Zoom in on the intersection regions for positive (b) and negative (c) wavevector $k_y$.}\label{fig5}
\end{figure}

To verify the theoretical results above, we have performed full micromagnetic simulations by MUMAX3 package \cite{mumax} (see Sec. IV \cite{SM} for simulation details). Figure \ref{fig5}(a) compares the magnon dispersion obtained from micromagnetic simulations (color map) with that from the linear spin-wave theory (dashed curve). One can clearly observe the level repulsion due to the very presence of the DDI. Upon zooming into the relevant crossing windows, we find that the magnon-magnon couplings are indeed asymmetric for positive and negative wavevectors [see Figs. \ref{fig5}(b) and (c)], which agrees with our theoretical predictions. One can also identify sizable discrepancy between analytical results and micromagnetic simulations. We attribute it to the omission of nonlinear magnon processes in the linear approximation and finite size effects in simulations. We point out that the asymmetric anticrossing induced by the DDI is quite robust against the spectral broadening caused by the Gilbert damping, high-order nonlinear magnon processes, and thermal fluctuations (Sec. V \cite{SM}). We therefore expect that the DDI-induced strong magnon-magnon coupling in altermagnets can be readily observed by real experiments employing the Brillouin light scattering (BLS) technique.
 
\textit{Discussion---}Besides the direct coupling induced by the DDI, an indirect coupling can be mediated by cavity photons. To demonstrate it, we consider a photon of circular polarization ${\bf e}_{\lambda}={\bf e}_x + i{\bf e}_z$, that propagates along the $y$ direction, and assume that the cavity field is uniform over the body of the altermagnet. Based on the same materials parameters, we can estimate the cavity-photon-induced magnon-magnon coupling $g_{\rm eff}\sim 0.01 \omega_{{\bf k},\pm}$, which is one order of magnitude weaker than the DDI-induced one (see Sec. VI \cite{SM} for details). We therefore conclude that, the cavity-mediated level repulsion between exchange magnons is embedded in the level broadening and can hardly be observed in real experiments. We also discuss the spin-flop phase when the applied field is large enough (Sec. VII \cite{SM}). It is found that the level-crossing disappears even without considering the DDI, recovering the results of conventional antiferromagnets.
 
 To summarize, we have predicted the DDI-induced strong coupling between short-wavelength magnons in altermagnets. We first showed that an in-plane magnetic field can only shift the magnon degeneracy from the origin to the exchange region but without destroying it. This feature is absent in conventional antiferromagnets. Moreover, we found that DDIs can effectively mix magnons with opposite chiralities, manifesting as a level repulsion in the magnon spectrum. The magnon-magnon coupling is shown to be highly anisotropic and strong enough that can be readily measured by practical experiments in $d-$wave altermagnets like $\rm KV_2Se_2O$ \cite{Jiang2025}, and $g-$wave altermagnets such as $\rm CrSb$ and $\rm MnTe$ \cite{Reimers2024,Lee2024}, using techniques like BLS. Our findings provide a new avenue for realizing strong and anisotropic magnon-magnon couplings, broadening the scope of exotic altermagnetic phenomena, and advancing the frontier of quantum magnonics.

\begin{acknowledgments}
This work was funded by the National Key R$\&$D Program under Contract No. 2022YFA1402802,  and the National Natural Science Foundation of China (NSFC) (Grants No. 12434003 and No. 12374103). Z. J. acknowledges the financial support from NSFC (Grant No. 12404125) and the China Postdoctoral Science Foundation (Grant No. 2024M750337).
\end{acknowledgments}

\end{document}